\documentclass[10pt,letterpaper]{article}
\usepackage[top=0.85in,left=1.0in,footskip=0.75in]{geometry}

\usepackage{amsmath,amssymb}

\usepackage{changepage}

\usepackage[utf8x]{inputenc}

\usepackage{textcomp,marvosym}

\usepackage{cite}

\usepackage{nameref,hyperref}

\usepackage[right]{lineno}

\usepackage{microtype}
\DisableLigatures[f]{encoding = *, family = * }

\usepackage[table]{xcolor}

\usepackage{array}

\newcolumntype{+}{!{\vrule width 2pt}}

\newlength\savedwidth

\raggedright
\usepackage[aboveskip=1pt,labelfont=bf,labelsep=period,justification=raggedright,singlelinecheck=off]{caption}

\makeatletter
\renewcommand{\@biblabel}[1]{\quad#1.}
\makeatother

\usepackage{lastpage,fancyhdr,graphicx}
\usepackage{epstopdf}
\fancyheadoffset[L]{2.25in}
\fancyfootoffset[L]{2.25in}
\usepackage{placeins} 

\newcommand{\fone}{F_{1}}
\newcommand{\pre}{\mathit{Pre}}
\newcommand{\rec}{\mathit{Rec}}
\newcommand{\jac}{\mathit{Jacc}}

\begin{document}
\vspace*{0.2in}

\begin{flushleft}
{\Large
\textbf\newline{Community evolution in retweet networks}
}
\newline
\\
Bojan Evkoski\textsuperscript{1,2},
Igor Mozeti\v{c}\textsuperscript{1},
Nikola Ljube\v{s}i\'{c}\textsuperscript{1},
Petra Kralj Novak\textsuperscript{1}
\\
\bigskip
\textbf{1} Department of Knowledge Technologies, Jozef Stefan Institute, Ljubljana, Slovenia
\\
\textbf{2} Jozef Stefan International Postgraduate School, Ljubljana, Slovenia
\\
\bigskip

* igor.mozetic@ijs.si

\end{flushleft}
\section*{Abstract}
Communities in social networks often reflect close social ties between their members and their evolution through time.
We propose an approach that tracks two aspects of community evolution in retweet networks: 
flow of the members in, out and between the communities, and their influence. 
We start with high resolution time windows, and then select several timepoints which exhibit 
large differences between the communities. For community detection, we propose a two-stage approach. 
In the first stage, we apply an enhanced Louvain algorithm, called Ensemble Louvain, 
to find stable communities. In the second stage, we form influence links between these communities, 
and identify linked super-communities. For the detected communities, we compute internal and external
influence, and for individual users, the retweet h-index influence.
We apply the proposed approach to three years of Twitter data of all Slovenian tweets. 
The analysis shows that the Slovenian tweetosphere is dominated by politics, 
that the left-leaning communities are larger, but that the right-leaning communities and users exhibit 
significantly higher impact. An interesting observation is that retweet networks change relatively 
gradually, despite such events as the emergence of the Covid-19 pandemic or the change of government.



\FloatBarrier
\section{Introduction}
\label{sec:intro}

With the ever-growing base of social media users, platforms such as Twitter are becoming a very valuable 
source of data for social analysis. Users on social media interact with each other, so it is natural to use graphs 
(where the users are nodes, and interaction between them are edges) to represent the structure of the user base. 
Nowadays, a lot of research in the field of complex networks is focused on social networks analysis. 
Due to the social media volatility, temporal analyses are needed for an in-depth understanding 
of the underlying phenomena. They can provide insights into the patterns and evolution of the social media 
landscape, and consequently to the society itself.

Change in the collective behaviour of groups in networks is referred to as community evolution~\cite{Dakiche2019tracking},
where communities in the networks are defined as groups of densely connected users. 
However, community detection methods are typically designed for static networks, 
and consequently have to be adapted for detecting changes in dynamic social media networks.

In our approach, we proceed by creating overlapping snapshots of the network through time,
and detect communities in each snapshot. We then track evolution of relevant communities over time. 
Several developments are needed to detect community evolution in terms of the flow of members 
in, out and between the communities, as well as to track the changes in the community influence.

We illustrate our approach to community evolution on a set of Slovenian tweets during the three year
period 2018--2020, roughly 13 million Twitter posts. 
Our initial research, where we performed a static community structure analysis of the data
showed strong polarization of the detected communities along the political dimension.
In the subsequent research, the basis of the current paper, we compared
community structures between different manually selected time windows~\cite{Evkoski2020polarization}. 
In the current paper, we describe a general set of techniques that enable 
semi-automatic analysis of the evolution of community structures and influence. 
These techniques make static community detection algorithms applicable to dynamic networks. 
We show a step-by-step application and insightful results of the proposed techniques 
on the Slovenian retweet networks.

\paragraph*{Related work.}
The temporal dimension is very valuable in modern analyses of complex networks. 
This has implications on how dynamic community discovery is designed and applied. 

The related approaches mostly depend on the representation of time. One can group them into 
three types: static/edge-weighted, snapshots, and temporal networks~\cite{Rossetti2018}. 
The first community discovery methods were applied to the so-called ``frozen in time'' networks, 
where the temporal dimension is not explicitly represented.
One operates with a single network (static or edge-weighted) that aggregates the whole period of interest.
This absence of the time dimension has two historical reasons: the graph theoretic origin of the field, 
and the scarcity of data at the time when the field of complex networks emerged~\cite{coscia2011classification}. 
Aggregation strategies have severe limitations as they cannot capture dynamics, 
hence are not suitable for dynamic community detection. 
Consequently, the second representation emerged---temporally ordered series of network snapshots. 
This approach allows for efficient tracking of changes in the network structure, 
thus increasing the expressiveness of the models, but at a cost of higher analytical complexity~\cite{Rossetti2018}. 
Finally, temporal networks were proposed that allow for a complete 
and fine-grained description of the network dynamics~\cite{holme2012temporal}.
The field of temporal network analysis is still under active development.
Explicit temporal network representation is rarely used for dynamic community discovery, 
as it considerably increases the complexity of the models, and cannot easily make use of
the existing community detection algorithms.

The snapshot approach crucially depends on the representation of time in the network relations. 
Two main scenarios were considered so far: perfect memory networks (also known as accumulative growth scenario),
and limited memory networks. Perfect memory permits only aggregation of nodes and edges, 
where the old nodes/edges cannot disappear. The limited memory scenario allows for nodes/edges to disappear over time. 
This is suitable in social network analysis, where the edge disappearance could indicate the decay of social ties.
The limited memory networks are implemented with various methods, including static, sliding, 
or dynamic-sized time windows, each method with its own strengths and weaknesses. 
In subsection~\nameref{sec:retworks} we propose a combined method that circumvents 
the drawbacks of the existing strategies by building a weight-decaying sliding time window network.
We then apply a snapshot selection method, described in subsection~\nameref{sec:timepoints}, 
where fixed static time windows, that contain 
the most information about the dynamics of the communities, are selected.

The second significant factor in dynamic community evolution is the way community detection 
is applied to the network 
snapshots~\cite{aynaud2013communities, hartmann2016clustering, lambiotte2016guide, Dakiche2019tracking, Rossetti2018}. 
Most of the existing approaches consider the following question: 
How do detected communities from one snapshot affect other snapshots
(usually future-adjacent)? There are three groups of approaches: 
non-evolutionary, evolutionary, and coupling. The first one, also known as instant-optimal 
or two-stage approach, considers that communities already existing at time $t$ depend only on the current state 
of the network at time $t$. A two-stage approach first detects communities at each snapshot,
and then matches the detected communities~\cite{chen2010detecting, bota2010community}. 
The obvious drawback of this approach is that the knowledge gained about the communities at snapshot $t$-1 is 
not used for communities at snapshot $t$. Yet, our method shows that this is not necessarily
a weakness, when one is interested in detecting maximal changes in the community structure. 
In the evolutionary approach, also known as temporal trade-off, communities at snapshot $t$ 
do not only depend on the network at the same time $t$, but also on the past evolution 
of the network~\cite{alvari2014community, agarwal2012real, crane2015community}. 
The coupling approach shifts the focus from detecting communities at snapshot $t$, 
to community detection considering pairs of adjacent snapshots, or even to the whole network 
evolution~\cite{aynaud2011multi, gauvin2014detecting}.

Although there is a plethora of approaches, with all their advantages and drawbacks, 
most of the methods suffer from a common issue---the instability of community detection 
algorithms~\cite{aynaud2010static}. Community detection algorithms have different weaknesses, 
but the instability of the results is their common issue in the temporal scenarios. 
This is specially problematic in the evolutionary approach to dynamic community detection 
since the local instability also affects the time dependent communities. 
In other words, a ``bad'' run of the community detection algorithm
influences the results of detection at the subsequent snapshots. 
This instability is also an issue for the community evolution analysis in our work, 
as one cannot distinguish if the community differences are due to the real-world events reflected 
in the dynamic complex network, or are they simply a consequence of the instability of the algorithm. 
To address this issue, 
we propose an Ensemble Louvain algorithm which to some extend solves the instability of the well-known 
Louvain algorithm for community detection.

\paragraph*{Structure of the paper.}
The main body of the paper is in the \nameref{sec:results} section.
We start with a brief overview of the data collected in the \nameref{sec:twitter} subsection.
In the \nameref{sec:retworks} subsection we describe how the network snapshots are created.
Network partitions, generated by an extension of the Louvain algorithm,
are described in the \nameref{sec:community} subsection.
Evolving communities in adjacent partitions are compared in the \nameref{sec:fone} subsection.
In \nameref{sec:timepoints} we show how to select just a few relevant timepoints
out of the whole timeline sequence.
Two types of transitions are depicted in the \nameref{sec:sankey} subsection.
We then define internal and external influence in the \nameref{sec:super} subsection.
In the last subsection \nameref{sec:hindex} we show the most influential users
in our dataset.
In \nameref{sec:concl} we wrap up our approach to community evolution and present
main plans for future research.
The \nameref{sec:methods} section provides some additional details.
The \nameref{sec:data} subsection describes a specialized tool used for Twitter acquisition.
\nameref{sec:ensemble} gives details of the community detection algorithm applied,
and some preliminary evaluation results.
The last subsection on \nameref{sec:bcubed} defines the measures used throughout the paper.

\FloatBarrier
\section{Results}
\label{sec:results}

\FloatBarrier
\subsection{Twitter data}
\label{sec:twitter}

Social media, and Twitter in particular, are widely used to study various social phenomena.
For this study, we collected a set of all Slovenian tweets in the three year
period, from January 1, 2018 until December 28, 2020. The set of almost 13 million tweets 
represents an exhaustive collection of Twitter activities in Slovenia. 
See the~\nameref{sec:methods} section for details of the Twitter data acquisition.

\begin{figure*}[!ht]
\begin{center}
\hspace*{-0.5in}
\includegraphics[width=19cm]{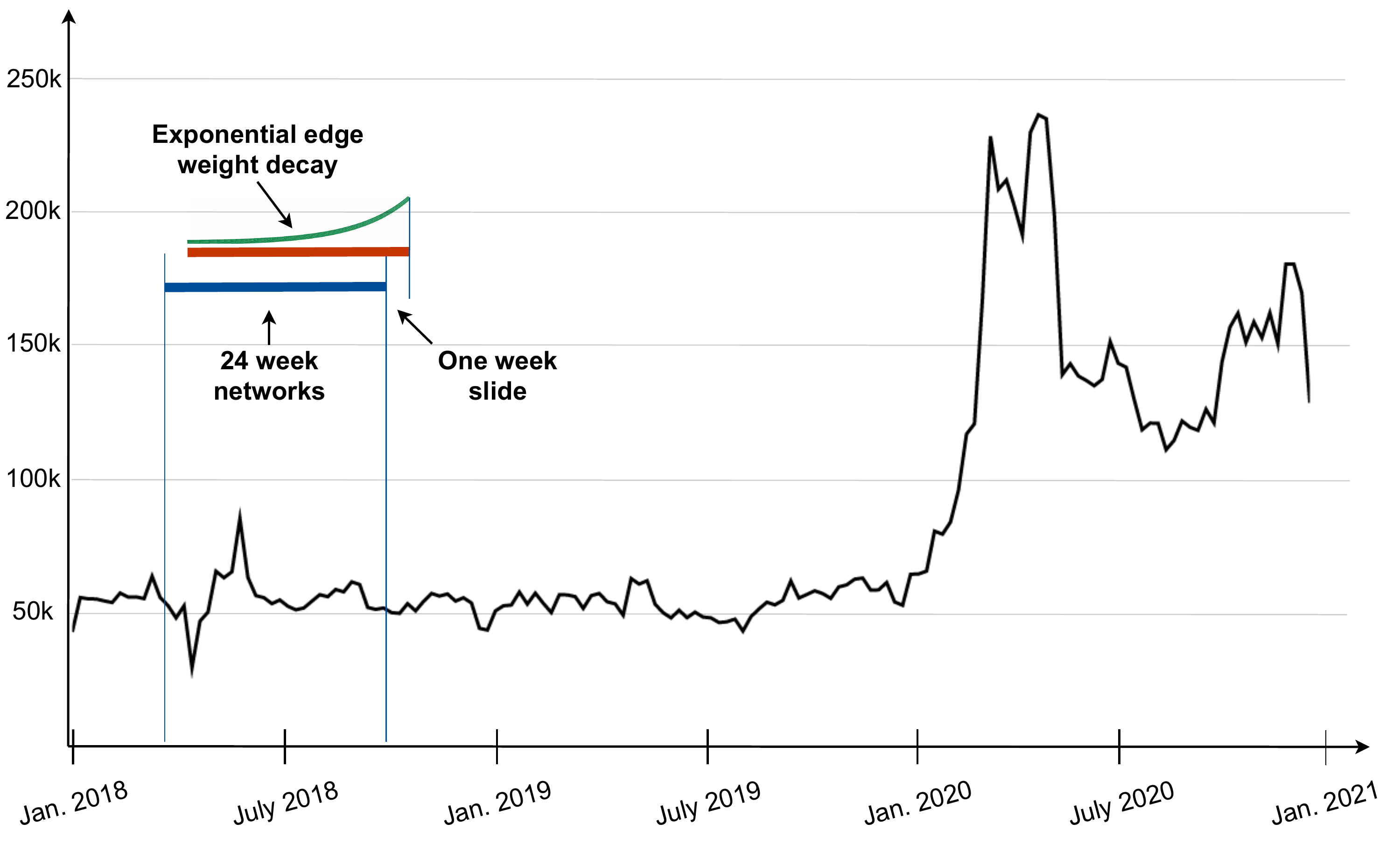}
\caption{
\textbf{Weekly volume of Slovenian Twitter data, collected over the three year period.}
The retweet network observation window is 24 weeks (blue and red lines), 
with exponential weight decay (half-time of 4 weeks, green curve), 
and one week sliding window (difference between the red and blue line).
Note a large increase of Twitter activities at the emergence of the Covid-19 pandemic,
which also coincided with the change of the left-wing to the right-wing
government in Slovenia.
}
\label{fig:fig1}
\end{center}
\end{figure*}

Fig~\ref{fig:fig1} shows the weekly volume of tweets collected during the three years. The number of
tweets is fairly stable, around 50,000 per week, until the emergence of the Covid-19 pandemic in
March 2020. At this point, we observe a four-fold increase of Twitter activities. This also coincides 
with the change of government in Slovenia, from the left-wing to the right-wing.
A minor peak can also be observed around June 2018, at the time of the snap parliamentary elections.
It turns out that most of the tweets are related to politics and ideology, and, after March 2020, to
policies concerning the handling of the pandemic. The following is a list of the most important 
political events in Slovenia during the 2018-2020 period. A prime minister (PM) at the time, 
with his Twitter handle, is given in parentheses:
\begin{itemize}
\item March 14, 2018 - left-wing government resignation (PM @MiroCerar),
\item June 8, 2018 - snap parliamentary elections,
\item September 13, 2018 - new left-wing government formation (PM @sarecmarjan),
\item January 27, 2020 - left-wing government resignation (PM @sarecmarjan),
\item March 13, 2020 - right-wing government formation (PM @JJansaSDS),
\item March, 2020 - emergence of the Covid-19 pandemic in Slovenia.
\end{itemize}

\FloatBarrier
\subsection{Retweet networks}
\label{sec:retworks}

Twitter provides different forms of interactions between the users: follows, mentions, replies, 
and retweets. The most useful indicator of social ties between the Twitter users are retweets. 
When a user retweets a post, it is distributed to all of its followers, just as if it were 
an originally authored post. Users retweet content that they find interesting or agreeable.
Despite the fact that it does not always signify an endorsement (e.g., tweets by the former
U.S. president Trump),
in large number of cases retweets indicate links between the like-minded users. In particular, 
in politics retweets very well reflect the actual political alignments and influence.
For example, it was demonstrated that political parties and nationalities of the members of the
European Parliament can be reconstructed solely from their retweet activities~\cite{Cherepnalkoski2016retweet}.
There is also a correspondence between the co-voting and retweeting in the European Parliament, 
while higher Twitter activity was observed for the right-wing parties~\cite{Cherepnalkoski2016cohesion}.
In the case of Brexit, the Leave proponents showed much higher activity and influence 
on Twitter than the Remain proponents~\cite{Grcar2017stance}.

A retweet network is a directed graph $G$. The nodes are Twitter users and the edges are retweet 
links between the users. An edge is directed from the user $A$ who posts a tweet to the user $B$ 
who retweets it. The edge weight is the number of retweets posted by $A$ and retweeted by $B$. 
For the whole three year period of Slovenian tweets, there are in total 18,821 users (nodes) 
and 4,597,865 retweets (sum of all weighted edges).

To study dynamics of the retweet networks, we form several network snapshots from our Twitter data.
In particular, we select a network observation window of 24 weeks (about six months),
with a sliding window of one week. This provides a relatively high temporal resolution between
subsequent networks, but later we show how to select the most relevant intermediate timepoints
(see subsection~\nameref{sec:timepoints}). Additionally, we employ an exponential edge weight decay,
with half-time of 4 weeks (see Fig~\ref{fig:fig1}). The reason for this temporal weight decay
is to eliminate the effects of the trailing end of the moving network snapshots.

The set of network snapshots thus consists of 133 overlapping observation windows, with temporal delay of one week. 
The snapshots start with network $G_0$ (January 1, 2018 - June 18, 2018) and end with network 
$G_{132}$ (July 13, 2020 - December 28, 2020).

\FloatBarrier
\subsection{Community detection}
\label{sec:community}

Informally, a network community is a subset of nodes more densely linked between themselves 
than with the nodes outside the community.
There are several formal definitions of communities and different methods to detect them.
A practical review that provides strengths and weaknesses of the most popular methods
is provided in~\cite{Fortunato2016community}.

A standard community detection method is the Louvain algorithm~\cite{Blondel2008fast}.
Louvain finds a partitioning of the network into communities, such that the modularity
of the partition is maximized. For a partition, the modularity measures the density 
and structure of its communities: the fraction of edges within the communities, 
as compared to the expected fraction of randomly distributed edges in the network~\cite{Newman2006modularity}.
The Louvain algorithm is computationally efficient, well suited for large networks,
and does not require ex-ante assumptions about the number or size of the
communities~\cite{Lancichinetti2009community}.

However, there are several problems with the modularity maximization~\cite{Fortunato2016community}.
One, from a theoretical point of view, is that there are typically exponentially many 
distinct partitions whose modularity scores are very close to the global 
maximum~\cite{Good2010modularity}. As a consequence, from a practical point of view,
the Louvain algorithm yields different partitions for different trials on the same network
(see Fig~\ref{fig:fig7} in~\nameref{sec:methods} for an example).

We address this instability problem of Louvain by applying the Ensemble Louvain algorithm.
We run 100 trials of Louvain and compose communities with nodes that co-occur in the same
community above a given threshold, 90\% of the trials in our case. This results in 
relatively stable communities of approximately the same size as produced by individual
Louvain trials. Details of the Ensemble Louvain algorithm are in the~\nameref{sec:methods} section.

Our 133 retweet network snapshots are directed graphs, $G_0, \ldots, G_{132}$, with weighted edges.
For community detection, we transform them into undirected graphs. When a pair of nodes is
linked with two weighted edges of the opposite direction, we create an undirected edge with the
sum of the original edge weights. When a pair of nodes is linked with a single directed edge, 
we simply drop the direction. We then run the Ensemble Louvain on all the 133 undirected 
network snapshots, resulting in 133 network partitions, $P_0, \ldots, P_{132}$.

\FloatBarrier
\subsection{Measuring community similarity}
\label{sec:fone}

The sequence of network partitions, $P_0, \ldots, P_{132}$, produced by Ensemble Louvain, varies.
Community structure changes, new nodes join some communities, and some nodes disappear
from a network snapshot. To study community evolution, one has to compare subsequent
network partitions. 

There are several measures to evaluate network communities,
in particular in relation to the ``ground truth''. Two widely used measures are
Adjusted Rand Index (ARI)~\cite{Hubert1985ARI} and
Normalized Mutual Information (NMI)~\cite{Danon2005NMI}.
In this study we use the BCubed measure, extensively evaluated in the context
of clustering~\cite{Amigo2009bcubed}. BCubed yields evaluation results similar
to ARI and NMI (see Fig~\ref{fig:fig7} in~\nameref{sec:methods}).
However, there are several advantages of BCubed, useful in the context of community evolution.
In particular, we extend the BCubed measure to account for the new and lost nodes 
between two network partitions.

BCubed decomposes evaluation into calculation of precision and recall of each node in the network. 
The precision ($\pre$) and recall ($\rec$) are then combined into the $\fone$ score,
the harmonic mean:
$$
\fone = 2\,\frac{\pre \cdot \rec}{\pre + \rec}.
$$
Details of computing $\pre$ and $\rec$ for individual nodes, communities
and network partitions are in the~\nameref{sec:methods} section. Here we just emphasize
that our extended BCubed $\fone$ is different and more general than the $\fone$ score 
proposed by Rossetti~\cite{Rossetti2016f1}.

In the following, we refer to our extended BCubed $\fone$ score as simply $\fone$.
When we compare two network partitions, $P_{t}$ and $P_{t-1}$, we consider a
partition earlier in time $P_{t-1}$ as ``ground truth'', and evaluate the
subsequent partition $P_{t}$ with respect to the previous one.
We write $\fone(P_{t} | P_{t-1})$ to denote the similarity of $P_{t}$ to $P_{t-1}$.
$\fone$ ranges from $0$ to $1$, where increasing $\fone$ indicates higher similarity
between the two partitions.

There are two special cases of $\fone$. When the two partitions consist of the same nodes,
just distributed differently between the communities, $\fone$ degenerates into $\textit{core-}\fone$.
$\textit{core-}\fone$ is directly compatible to ARI and NMI.
When the two partitions differ in the constituent nodes, i.e., there are new and lost nodes,
one can compute the theoretical maximum similarity, $\textit{max-}\fone$, where all the
nodes common to both partitions (the intersection) are assumed to be in one community.
$\textit{max-}\fone$ thus measures similarity of two sets and is directly
related to the Jaccard index. See~\nameref{sec:methods} for details.

Fig~\ref{fig:fig2} (red line) shows pairwise $\fone(P_{t} | P_{t-1})$ differences between the
retweet network partitions at weekly timepoints $t=1,2,\ldots,132$.
The $\fone$ scores are relatively high, typically in the range $[0.8, 0.9]$.
The largest negative peak, indicating the highest dissimilarity,
$\fone(P_{92} | P_{91}) = 0.74$, occurs between the partitions which end on
March 16 and 23, 2020, respectively. These dates closely follow the change
of government in Slovenia and first policy reactions to the emergence of the Covid-19 pandemic.

\FloatBarrier
\subsection{Selection of timepoints}
\label{sec:timepoints}

The weekly differences between the network partitions are relatively low.
The retweet network communities apparently do not change drastically at this
relatively high time resolution. Moving to lower time resolution means
choosing timepoints which are further apart, and where the network communities
exhibit more pronounced differences.

We formulate the timepoint selection task as follows. 
Let assume that the initial and final timepoints are fixed, 
corresponding to the partitions $P_0$ and $P_n$, respectively.
For a given $k$, select $k$ intermediate timepoints such that the differences
between the corresponding partitions are maximized, i.e., the $\fone$
scores are minimized:
$$
min \left( \sum_{i=1}^{k} \fone(P_i | P_{i-1}) + \fone(P_n | P_k) \right).
$$
There are $\binom{n-1}{k} \cdot {k!}$ possible selections of timepoints, i.e.,
the number of selections grows exponentially with $k$.
We therefore propose a simple heuristic algorithm which finds $k$ approximate timepoints.
The algorithm works top-down and starts with the full, high resolution timeline with
$n+1$ timepoints, $t=0,1,\ldots,n$ and corresponding partitions $P_{t}$.
At each step, it finds a triplet of adjacent partitions $P_{t-1}, P_{t}, P_{t+1}$ 
with minimal differences:
$$
max \left( \fone(P_t | P_{t-1}) + \fone(P_{t+1} | P_t) \right)
$$
and eliminates $P_t$ from the timeline. At the next step, the difference
$\fone(P_{t+1} | P_{t-1})$ fills the gap of the eliminated timepoint $P_t$.
The algorithm thus finds the $k$ (non-optimal) timepoints in $n-1-k$ steps.
While efficient, this approach to the relevant timepoint selection is not suitable for incremental, 
stream-based network processing since it assumes that the final timepoint is fixed.

\begin{figure*}[!ht]
\begin{center}
\includegraphics[width=\textwidth]{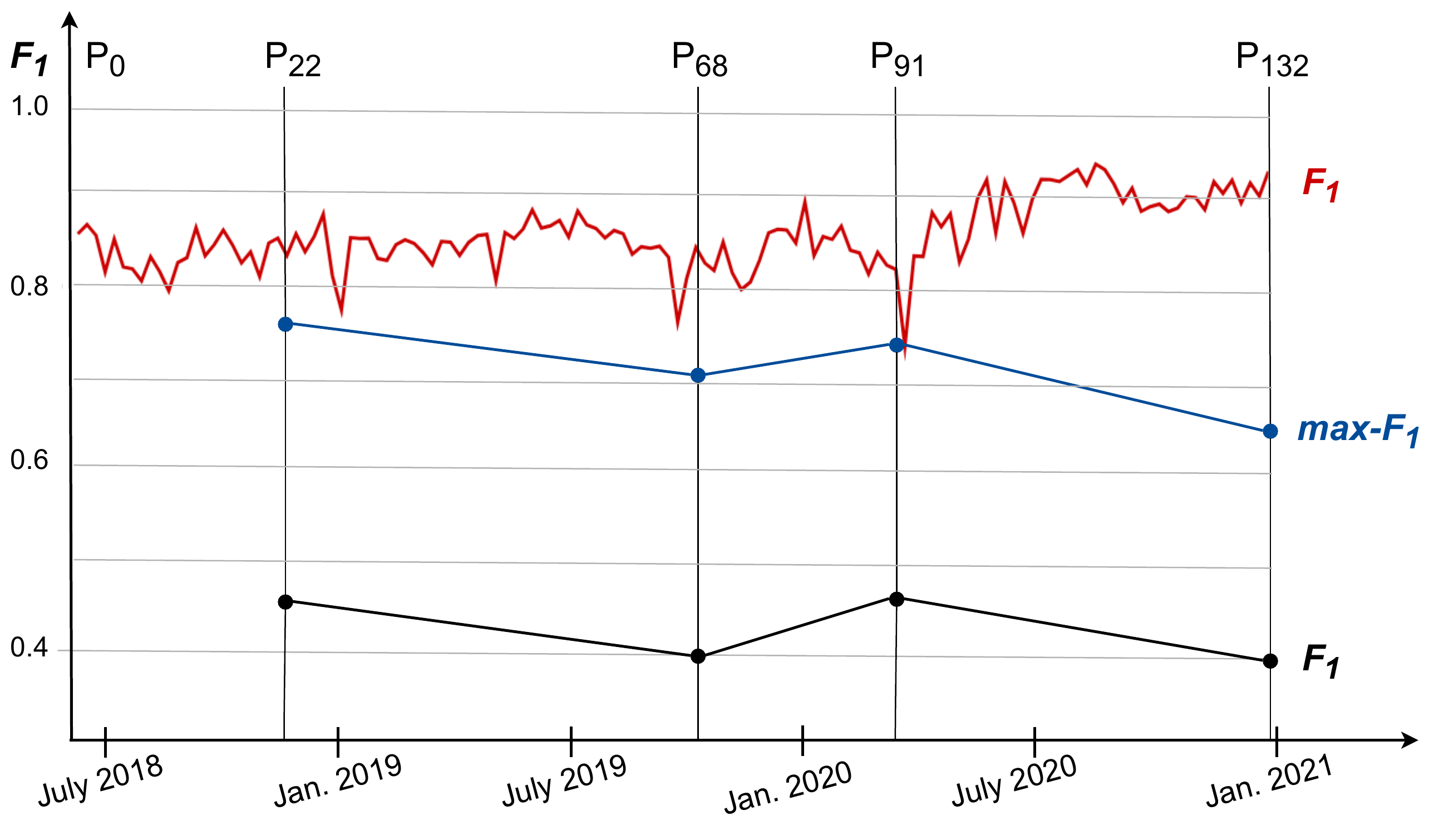}
\caption{
\textbf{Differences between the adjacent network partitions measured by the $\fone$ score.}
The red line at the top shows weekly differences $\fone(P_{t} | P_{t-1})$ at timepoints $t=1,2,\ldots,132$.
The five selected partitions are denoted by $P_{0}, P_{22}, \ldots, P_{132}$.
The middle blue line shows the theoretical maximum $\textit{max-}\fone$ differences 
between distant partitions at the selected timepoints $t=0, 22, 68, 91, 132$.
The bottom black line shows the standard $\fone$ differences.
}
\label{fig:fig2}
\end{center}
\end{figure*}

For our retweet networks, we experimented with several values of $k$ and
eventually settled with $k$=3 which provides much lower, but still meaningful time resolution.
This resulted in the selection of the following network partitions: $P_{0}, P_{22}, P_{68}, P_{91}, P_{132}$.
Fig~\ref{fig:fig2} shows the $\fone$ differences (black line) between the adjacent partitions.

The selected timepoints are on average 26 weeks apart, varying between five and ten months.
The differences between the network partitions are increasing with temporal
distance, but are still relatively uniform, $\fone$ is in the range $[0.4, 0.5]$.
Due to these small differences, the timepoint selection procedure is not very robust.
The selected timepoints should be considered approximate and can vary for several
weeks in both directions. As a consequence, the selected timepoints should not be
interpreted as indicators of specific events at specific dates, but should rather 
help in understanding longer terms qualitative transitions in community evolution.

Fig~\ref{fig:fig2} also shows the theoretical maximum differences $\textit{max-}\fone$ (blue line), 
where it is assumed that all the common nodes in two adjacent partitions are in one community,
and only the intersection size and the number of new and lost nodes affect the score.
The $\textit{max-}\fone$ scores, dropping from $0.77$ to $0.63$, show increasing
fluctuation of nodes in and out of the partitions. In the next subsection we show 
two visualizations of transitions between these five network partitions.

\FloatBarrier
\subsection{Visualization of community transitions}
\label{sec:sankey}

We present two visualizations of transitions between selected network partitions
as Sankey diagrams. A Sankey diagram is a type of flow diagram in which the width of the 
bands is proportional to the flow rate.

\begin{figure*}[!ht]
\begin{center}
\hspace*{-0.5in}
\includegraphics[width=19cm]{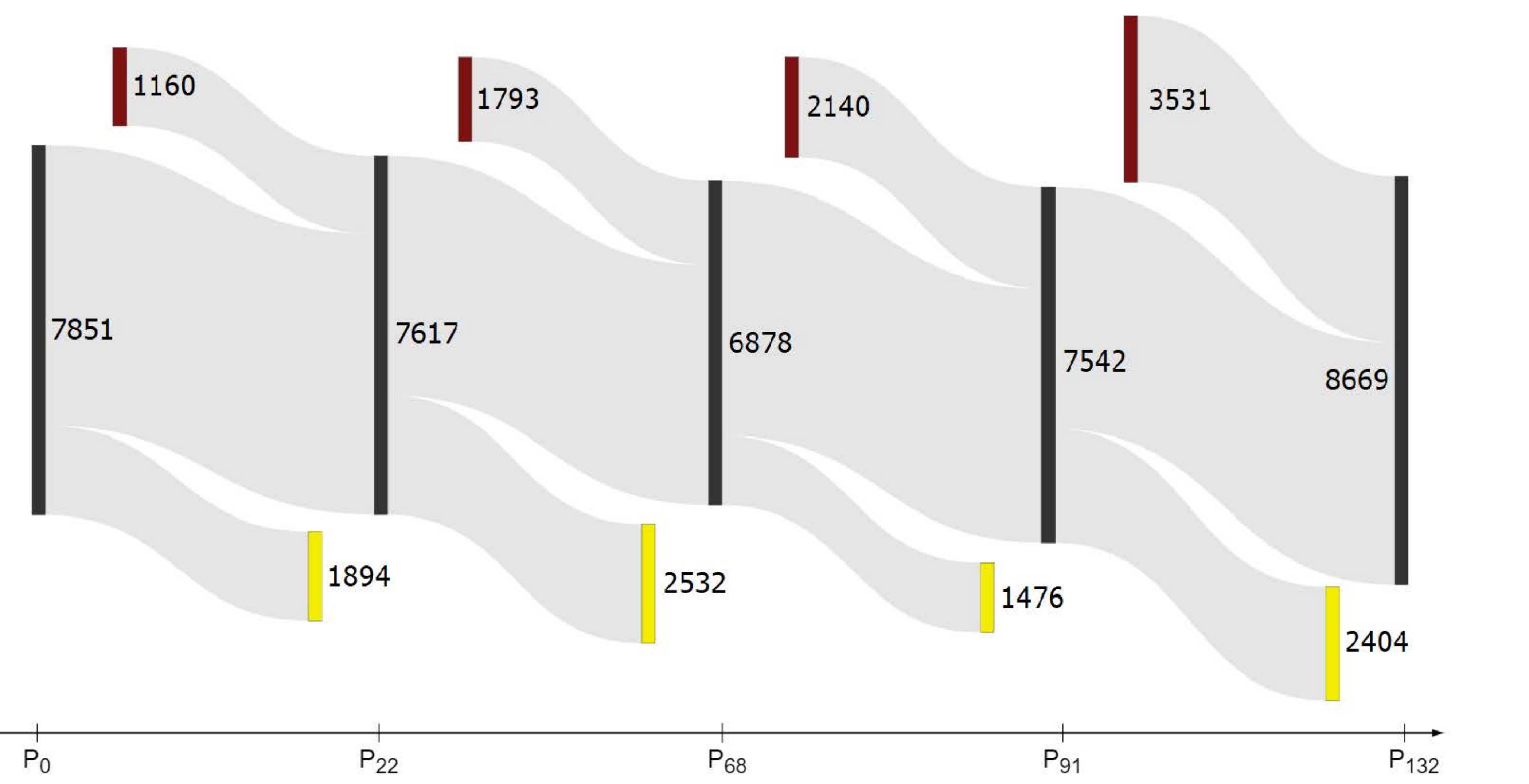}
\caption{
\textbf{A Sankey diagram showing major transitions between the five selected timepoints
$P_{0}, P_{22}, \ldots, P_{132}$.}
The numbers indicate core nodes (black), new nodes (brown, at top), 
and lost nodes (yellow, at bottom) between two adjacent network partitions. 
The differences between the adjacent partitions are quantified
by $\textit{max-}\fone(P_{i} | P_{i-1})$, shown with blue line in Fig~\ref{fig:fig2}.
Note a relatively large in- and out-flow of new and lost nodes between the partitions.
}
\label{fig:fig3}
\end{center}
\end{figure*}

In Fig~\ref{fig:fig3} we show inflows of new nodes, outflows of lost nodes, and
transition flows of core (intersection) nodes between the selected network partitions.
Note that only about half of the nodes remain in the core transitions.
Therefore it is crucial that the community similarity measure takes new and
lost nodes into account. This diagram ignores the internal community structure,
and corresponds to the theoretical maximum $\textit{max-}\fone$ (shown in 
Fig~\ref{fig:fig2}) where all the core nodes are assumed to be in the same community.

\begin{figure*}[!ht]
\begin{center}
\hspace*{-0.5in}
\includegraphics[width=19cm]{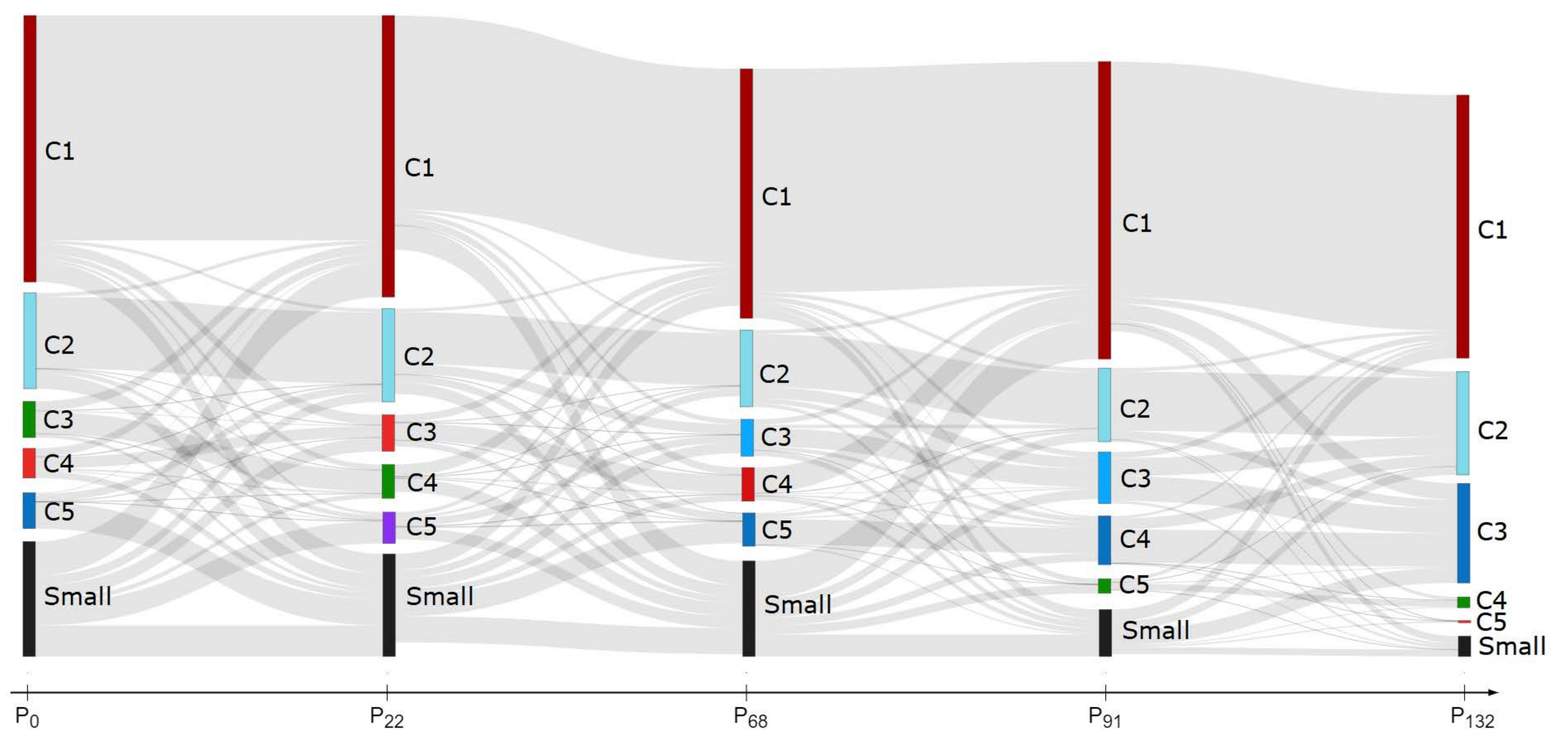}
\caption{
\textbf{A Sankey diagram showing transitions between the five largest communities 
$C1, \ldots, C5$ at the selected timepoints $P_{0}, P_{22}, \ldots, P_{132}$.}
The remaining smaller communities are 
labeled as Small, and new and lost nodes are not shown here. The differences between the
adjacent partitions are quantified by $\fone( P_{i} | P_{i-1})$, shown with black line
in Fig~\ref{fig:fig2}. The left-leaning communities are in shades of red, the right-leaning
communities are in shades of blue, and the Sports community is green.
}
\label{fig:fig4}
\end{center}
\end{figure*}

Fig~\ref{fig:fig4} is more detailed and shows the internal community structure of the cores,
with the top five communities $C1, \ldots, C5$ at each selected timepoint. 
All the remaining smaller communities are appended together into a single Small community.

The top communities were manually scanned for the most influential users (see subsection
\nameref{sec:hindex}) and discussion topics. 
It turns out that most of the communities are structured around political figures 
(either politicians, public figures, or journalists with clear political
orientation) \cite{Evkoski2021commhate}, 
and that political and ideological topics are prevailing~\cite{Kralj2021hate}.
Thus, the top communities can be classified into three categories:
left-leaning (most influential users are part of the left-wing structures),
right-leaning (most influential users are part of the right-wing), and
Sports (users and topics are clearly related to sports).
In Fig~\ref{fig:fig4}, the left-leaning communities are in shades of red and the right-leaning 
communities are in shades of blue. The only non-political community is Sports, in green, 
represented by the following sequence of communities:
$$
C3_{0} \mapsto C4_{22} \mapsto (C \subset Small)_{68} \mapsto C5_{91} \mapsto C4_{132}.
$$
A community $Ci$ at timepoints $t$ is denoted by $Ci_{t}$.
Note that at timepoint $t$=68 the Sports community is absorbed into the Small community.

The political communities are considerably larger than Sports.
Let us first consider some right-leaning communities, which feature the current Slovenian 
prime minister of the right-wing government, with a Twitter handle @JJansaSDS. 
He was initially a member of relatively small communities
that at timepoint $t$=22 did not even make it into the top five:
$$
C5_{0} \mapsto (C \subset Small)_{22} \mapsto C5_{68} \mapsto C4_{91} \mapsto C3_{132}.
$$
Only after the right-wing government took over in March 2020 (timepoints $t$=91, 132)
did his community grow considerably.

On the political left-wing, there is the $C1$ community that grows and shrinks with
time, but remains by far the largest community throughout the three year period.
The former Slovenian prime minister
(between September 2018 and March 2020), with a Twitter handle @sarecmarjan,
was a member of $C1$ for most of the time:
$$
C1_{0} \mapsto C1_{22} \mapsto C4_{68} \mapsto C1_{91} \mapsto C1_{132}.
$$
Only in the second half of his government ($t$=68) did he feature prominently 
in his own community $C4$. The left-wing Slovenian prime minister before him
(until March 2018), with a Twitter handle @MiroCerar, was initially 
a member of smaller communities on the left-wing, and recently joined $C1$:
$$
C4_{0} \mapsto C3_{22} \mapsto C4_{68} \mapsto C1_{91} \mapsto C1_{132}.
$$

It is interesting to observe the official Slovenian government Twitter account @vladaRS.
It moves from the left-leaning to the right-leaning communities as the left-wing government is
replaced by the right-wing one, but with some delay:
$$
C4_{0} \mapsto C3_{22} \mapsto C4_{68} \mapsto C1_{91} \mapsto C3_{132}.
$$
@vladaRS matches the @MiroCerar community at $t$=0, 22, the @sarecmarjan community at $t$=68, 91,
and the @JJansaSDS community at $t$=132. This is another piece of evidence that retweet
communities evolve gradually and that it takes a while before events with a high
impact are reflected in a new community structure.

To further characterize political polarization and community evolution, 
we now turn attention from the community membership to the retweet links between the communities.

\FloatBarrier
\subsection{Identification of super-communities}
\label{sec:super}

Twitter users differ in how prolific they are in posting tweets, and in the impact these
tweets make on the other users. One way to estimate the influence of a Twitter user
is to consider how often are its posts retweeted. Similarly, the influence of a
community can be estimated by the total number of retweets of their posts.
Retweets within the community indicate internal influence, and retweets outside of
the community indicate external influence. This approach to characterize
influential users and communities was already applied to a wide
range of environmental issues discussed on Twitter~\cite{Sluban2015sentiment}.

\begin{figure*}[!ht]
\begin{center}
\hspace*{-0.5in}
\includegraphics[width=19cm]{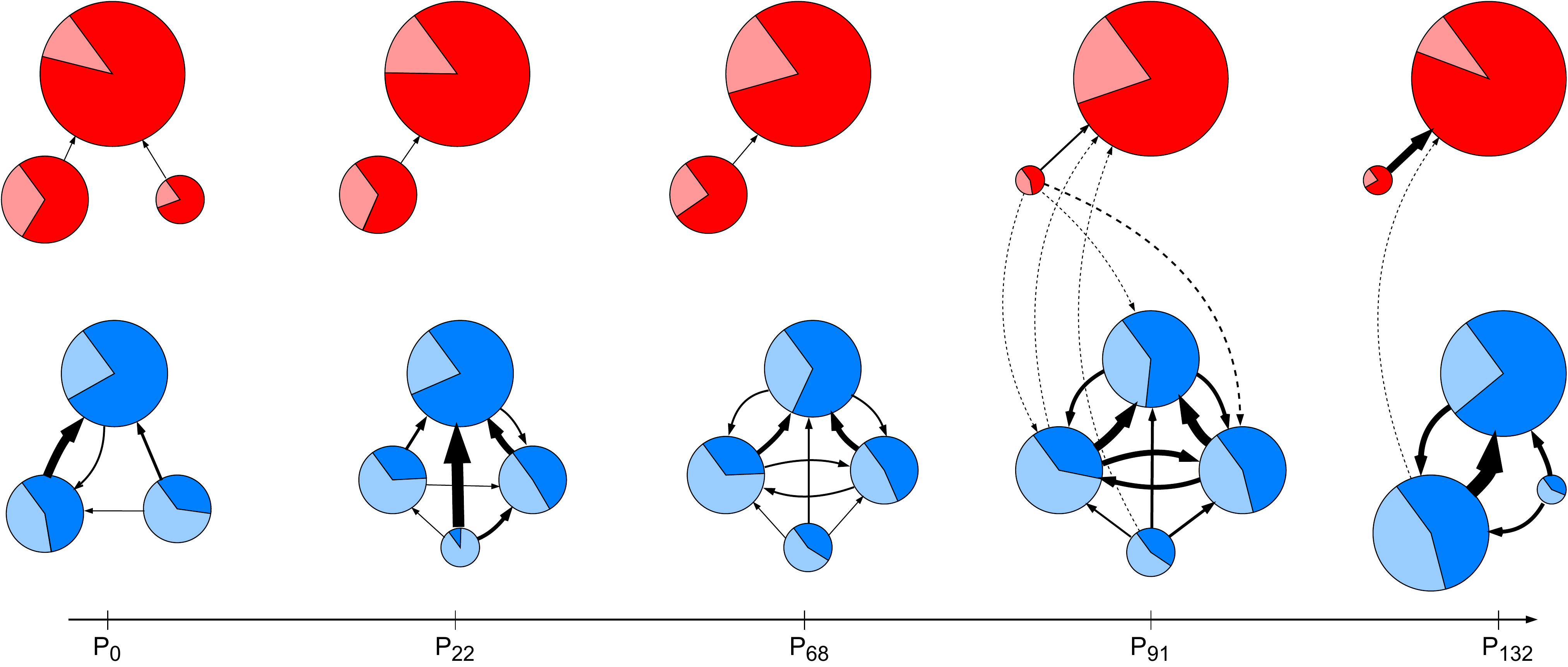}
\caption{
\textbf{Identification of super-communities from the meta-networks.}
Nodes are detected communities $C1, \ldots, C7$ at different timepoints, 
and edges denote average external influences.
A node diameter is proportional (cube-root) to the number of community members,
darker area corresponds to the internal influence, and lighter area to the
external influence. 
Red communities are part of the Left-wing super-community, blue communities
are part of the Right-wing super-community, and the remaining Sports community
is not shown. Dashed edges show rare and relatively
weak links between the Left- and Right-wing communities.
}
\label{fig:fig5}
\end{center}
\end{figure*}

In this subsection we focus on community influence and subsequent identification of
super-communities. Another measure of individual influence is described
in the next subsection~\nameref{sec:hindex}. In our retweet networks,
the number of retweets is represented by the weighted out-degree of a node.
Let $W_{ij}$ denote the sum of all weighted edges between communities $C_i$ and $C_j$.
The average community influence $I$ is defined as:
$$
I(C_i) = \frac{\sum_{j} W_{ij}}{|C_i|},
$$
i.e., the weighted out-degree of $C_i$, normalized by its size.
The influence $I$ consists of the internal $I_{int}$ and external $I_{ext}$ component,
$I = I_{int} + I_{ext}$, where
$$
I_{int}(C_i) = \frac{W_{ii}}{|C_i|},
$$
and
$$
I_{ext}(C_i, C_j) = \frac{\sum_{i \neq j} W_{ij}}{|C_i|}.
$$
We compute internal and external influence of the retweet communities detected
at the selected timepoints $t=0, 22, 68, 91, 132$. The communities which are
politically left- or right-leaning are shown in Fig~\ref{fig:fig5}, the Sports community
is omitted. A community, proportional to its size, is depicted as a pie chart,
indicating its internal and external influence. A pair of communities $C_i, C_j$
is linked by a weighted directed edge from $C_i$ to $C_j$,
with the weight equal to the external influence $I_{ext}(C_i, C_j)$.

The meta-networks in Fig~\ref{fig:fig5} support clear identification of two super-communities:
Left-wing and Right-wing. A super-community exhibits relatively strong external influence links
between its constituent communities. However, there are considerable differences between
the Left-wing and Right-wing super-communities. The Left-wing is larger, and its
communities have higher internal influences. The Right-wing, on the other hand, 
has stronger inter-community links, its communities have higher external influences, 
and appears more cohesive. Note that there are barely any links between the Left-wing and 
Right-wing communities, a characteristics of echo chambers and political 
polarization~\cite{DelVicario2016echo}.

\begin{figure*}[!ht]
\begin{center}
\includegraphics[width=\textwidth]{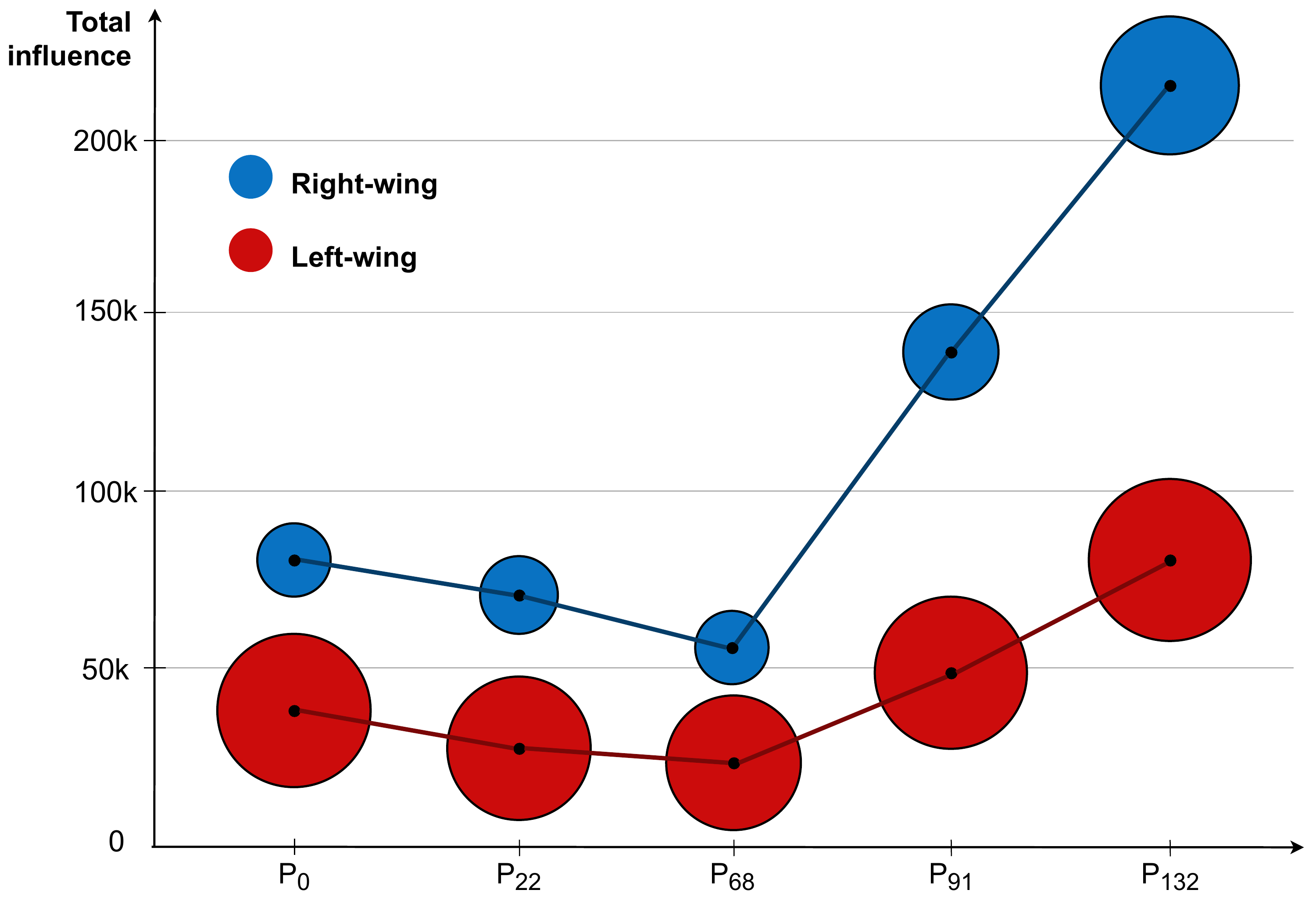}
\caption{
\textbf{Total weighted out-degree influence for both super-communities.}
A super-community size is proportional to the number of its members.
Total influence is the sum of weighted out-degree influences of all super-community members.
Note that the influence of the Right-wing super-community is at least twice as large as 
the influence of the Left-wing super-community and increasing with time, despite the fact that
it is considerably smaller.
}
\label{fig:fig6}
\end{center}
\end{figure*}

In Fig~\ref{fig:fig6} we show the total influence of both super-communities.
Total influence of a super-community is the sum of weighted out-degrees of all its members,
without normalization. The Right-wing super-community is typically half the size of
the Left-wing, approaching in size only at the last timepoint $t$=132.
However, the influence of the Right-wing is always considerably higher, with the gap even increasing
after the right-wing government took over in March 2020 (timepoints $t$=91, 132).

\FloatBarrier
\subsection{Retweet h-index influence}
\label{sec:hindex}

Weighted out-degree is a useful measure of influence for communities and
super-communities. However, we propose a different measure of influence for individual Twitter users.
The user influence is estimated by their retweet h-index, an adaptation of
the well known Hirsch index~\cite{Hirsch2005index} to Twitter.
The retweet h-index takes into account the number of tweets posted, as well as the impact
of individual posts in terms of retweets.

A user with an index of $h$ has posted $h$ tweets and each of them was retweeted at least $h$ times.
Let \textit{RT} be the function that returns the number of retweets for each original post.
The values of \textit{RT} are ordered in decreasing order, 
from the largest to the lowest, and $i$ indicates the ranking position in the ordered list.
The h-index is then computed as follows:
$$
\text{h-index}(\textit{RT}) = \max_{i} \min( \textit{RT}(i), i).
$$
To the best of our knowledge, the retweet h-index was first used on Twitter data 
in the context of Brexit, to measure the influence of the Leave and Remain proponents~\cite{Grcar2017stance}.
Later, this measure of influence was termed a retweet h-index~\cite{Gallagher2021SustainedOA},
a term we also adopt here.

We compute the h-index and the h-index rank for all the users on Slovenian Twitter during
the three year period. For each super-community, Left-wing and Right-wing, we show the top ten most
influential users by h-index, ordered by the h-index rank (Table~\ref{tab:tab1}).
The users are ranked for the overall three year period, but the h-index and relative ranks are
also provided for the selected timepoints $t=0, 22, 68, 91, 132$.
Two of the top Twitter users, @vladaRS and @ukclj, do not remain in the same super-community, 
but move from the Left-wing to the Right-wing as the government changed.

There is a large difference between the members of the Right-wing and Left-wing super-communities.
The Right-wing members consistently take the top h-index ranks, while the Left-wing members
barely make it into the top 100 h-index ranks. This reaffirms the super-community
influence results from the subsection~\nameref{sec:super}, and is consistent with our previous
results. In the case of the European Parliament, higher Twitter activity was observed 
for the right-wing parties~\cite{Cherepnalkoski2016cohesion}.
In the case of Brexit, the Leave proponents showed much higher activity and influence 
on Twitter than the Remain proponents~\cite{Grcar2017stance}.

\begin{table}[!ht]
\begin{adjustwidth}{-0.5in}{0in}
\centering
\caption{
\textbf{Top ten influential users from each super-community, ranked by the overall h-index.}
Left-to-Right denotes users which moved from the Left-wing
to the Right-wing super-community (@vladaRS: the official Slovenian government account, and @ukclj: 
University Medical Centre Ljubljana). The top users in each super-community are @JJansaSDS 
(current prime minister), and @sarecmarjan (former prime minister), respectively. 
Each user is assigned the h-index rank (h-rank), the h-index (h-ind) for the overall 
three year period and the five selected timepoints ($P_{0}, \ldots, P_{132}$),
and the overall unweighted out-degree (out-deg). Note that the top Left-wing influential users 
barely reach the h-index rank of top 100.
}
\setlength{\tabcolsep}{0.4em} 
\begin{tabular}{|l|rrr|rr|rr|rr|rr|rr|}
\hline
 & \multicolumn{3}{c|}{Overall} & \multicolumn{2}{c|}{$P_{0}$} & \multicolumn{2}{c|}{$P_{22}$} & \multicolumn{2}{c|}{$P_{68}$} & \multicolumn{2}{c|}{$P_{91}$} & \multicolumn{2}{c|}{$P_{132}$} \\
Twitter user             & h-rank & h-ind & out-deg & h-rank & h-ind & h-rank & h-ind & h-rank & h-ind & h-rank & h-ind & h-rank & h-ind \\
\hline \hline
\bf{Right-wing:} &      &     &	   & 	&     &	   &   	 &    &	    & 	 &     &     &     \\
@JJansaSDS       &      1 &   168 &    2621 &      1 &    93 &      1 &    92 &      1 &    79 &      1 &    93 &      1 &   140 \\
@BojanPozar      &      2 &   111 &    2465 &      5 &    56 &      3 &    55 &      2 &    60 &      2 &    69 &      2 &    96 \\
@JozeMozina      &      3 &    99 &    1724 &      7 &    47 &      4 &    53 &      5 &    47 &      6 &    50 &      6 &    60 \\
@LahovnikMatej   &      4 &    95 &    2169 &      6 &    54 &      5 &    49 &     10 &    40 &      5 &    51 &      4 &    77 \\
@aleshojs        &      5 &    92 &    1609 &     32 &    29 &     39 &    28 &     34 &    28 &     20 &    37 &      3 &    80 \\
@Libertarec      &      6 &    83 &    2228 &      4 &    58 &      2 &    56 &      7 &    44 &     12 &    46 &      5 &    62 \\
@bmz9453         &      7 &    79 &    1715 &     34 &    28 &      7 &    46 &     23 &    31 &      7 &    48 &     10 &    56 \\
@BrankoGrims1    &      8 &    76 &    1403 &     31 &    29 &     10 &    39 &      4 &    47 &      4 &    52 &     27 &    43 \\
@ZigaTurk        &      9 &    75 &    2273 &     30 &    30 &     25 &    33 &     25 &    31 &     21 &    36 &     13 &    55 \\
@mrevlje         &     10 &    75 &    1998 &     28 &    30 &     12 &    39 &      8 &    42 &      3 &    56 &     14 &    53 \\
\hline
\bf{Left-to-Right:} &      &     &	   & 	&     &	   &   	 &    &	    & 	 &     &     &     \\
@vladaRS         &     14 &    72 &    2287 &    446 &     9 &    586 &     8 &    489 &     8 &     52 &    26 &      9 &    57 \\
@ukclj           &     32 &    59 &    2170 &      / &     / &    545 &     8 &    254 &    11 &     61 &    24 &     34 &    41 \\
\hline
\bf{Left-wing:} 		&      &     &	   & 	&     &	   &   	 &    &	    & 	 &     &     &     \\
@sarecmarjan     &     97 &    41 &     889 &    203 &    13 &    443 &     9 &    490 &     8 &    200 &    14 &     96 &    28 \\
@necenzurirano\_ &    103 &    40 &     876 &      / &     / &      / &     / &      / &     / &    225 &    13 &     48 &    37 \\
@PreglArjan      &    131 &    37 &    1349 &    117 &    17 &     52 &    25 &    258 &    11 &    227 &    13 &     84 &    29 \\
@rjerala         &    133 &    37 &    1014 &   1550 &     4 &   1480 &     4 &    793 &     6 &    365 &    10 &     68 &    32 \\
@strankalevica   &    140 &    36 &     886 &    170 &    14 &    160 &    16 &    219 &    12 &    114 &    19 &    105 &    27 \\
@AlHarlamov      &    141 &    35 &    1060 &    274 &    11 &    432 &     9 &    348 &     9 &    563 &     8 &     95 &    28 \\
@LukaMesec       &    149 &    35 &     613 &   3621 &     1 &   4595 &     1 &   3869 &     1 &    267 &    12 &     83 &    29 \\
@Matej\_Klaric   &    182 &    32 &    1236 &    239 &    12 &    144 &    17 &    218 &    12 &    157 &    16 &    124 &    25 \\
@STA\_novice     &    186 &    32 &    2204 &    367 &    10 &    255 &    13 &    325 &    10 &    143 &    17 &    162 &    23 \\
@SpletnaMladina  &    246 &    28 &    1424 &    135 &    16 &    200 &    14 &    166 &    14 &    177 &    15 &    197 &    21 \\
\hline
\end{tabular}
\label{tab:tab1}
\end{adjustwidth}
\end{table}

The Right-wing and Left-wing super-communities are led by the current (@JJansaSDS) and former
(@sarecmarjan) prime minister of Slovenia, respectively. The other top members are
either politicians, journalists, or public figures active on Twitter.
In the Left-wing, there are some media account (@necenzurirano\_, @STA\_novice,
@SpletnaMladina), and a political party account (@strankalevica - `The Left').

The only two users in Table~\ref{tab:tab1} which are clearly unrelated to politics are @ukclj 
(University Medical Centre Ljubljana) and @rjerala (Roman Jerala, a biochemist 
from the National Institute of Chemistry). They reached the rank of top 100 influencers
only after the emergence of the Covid-19 pandemic (@ukclj at $t$=91, and @rjerala at $t$=132).
They post tweets about the medical issues, drugs and vaccines related to the pandemic.
There are other influential users, tracking and commenting on the pandemic development,
which emerged recently, but they did not yet make it into the overall top ten h-index list.

\FloatBarrier
\section{Conclusions}
\label{sec:concl}

Social media, and Twitter in particular, are a rich source of data that reflects social
relations between the users. In the paper we exploit a specific type of networks
where retweets are used as links between the users. We demonstrate that in the retweet networks
meaningful communities are formed. We show that retweet influence reveals important
differences between different communities as well as between individual Twitter users.
The main focus of the paper is on the evolution of communities and influence through time,
and we address several issues relevant for the field of dynamic networks.

One problem is the instability of detected static communities by a standard
community detection Louvain algorithm. We propose to run an ensemble of Louvain trials, 
and detect stable communities through frequent co-occurrence of nodes across the trials. 
Preliminary evaluations of the Ensemble Louvain algorithm on some benchmark 
networks with known ``ground truth'' communities show promising results, 
and this is certainly one of the directions that needs to be further explored in the future.

We study network evolution by taking several static network snapshots with a sliding window.
One has to decide on the window size and the temporal resolution between the snapshots.
We decided on the 24 weeks window size and an exponential edge weight decay, with
half-time of 4 weeks. The edge decay removes the effect of the trailing end of the window,
and thus makes the choice of the window size less relevant. The choice of the
half-time decay, on the other hand, is subject to experimentation, and depends on the volume of
Twitter data. The chosen sliding window of one week provides high temporal resolution,
but again \textbf{}the choice of this parameter is not crucial. We propose a temporal zoom-out to a
lower time resolution, by computationally efficient selection of more distant timepoints
where the network partitions exhibit larger differences. An analysis of how robust is this
selection and what are meaningful ranges of distant timepoints is required in the future.

We apply and extend a measure of community similarity BCubed, 
which was originally introduced to evaluate
quality of document clustering, but does not appear to be used in the field of complex networks.
The $\fone$ score can measure differences between network communities with only partially
overlapping set of nodes. This is essential for comparing retweet networks,
where new nodes keep appearing and disappearing from the network snapshots.
An additional nice property of $\fone$ is that it degenerates into a well-known
set comparison coefficient, directly related to the Jaccard index.

A specially interesting result of this research is clear identification of super-communities
from external influence links between the detected communities.
The exiting problem, worth addressing in the future, is how to design a multi-stage 
super-community detection algorithm. This seems relevant for retweet networks in particular,
where a standard community detection algorithm produces a large set of fractured communities.

There are two follow-up directions of the current research, already undertaken:
classification of tweets by the level of hate speech and detection of discussion 
topics~\cite{Kralj2021hate}, and attribution of the hate speech to the detected
communities and types of users~\cite{Evkoski2021commhate}. 
The results show that most of the hate speech has the form of offensive tweets,
and that over 60\% of them can be attributed to a single right-leaning community 
of moderate size.

We illustrate our approach on a well-defined set of Slovenian tweets, of reasonable size,
but not extremely large. Our next step is to apply the same approach on two different,
but somehow related sets of Croatian and Serbian tweets. This will reveal which parameters
need to be tuned to specific datasets, and what seem to be domain-invariant properties
and methods, applicable to a wide range of domains.

\FloatBarrier
\section{Methods}
\label{sec:methods}

\subsection{Data collection}
\label{sec:data}

The three years of comprehensive Slovenian Twitter data cover the period from January~1, 2018 until December~28, 2020.
In total, 12,961,136 tweets were collected.
We used the TweetCaT tool~\cite{Ljubesic2014tweetcat} for Twitter data acquisition.

The TweetCaT tool is specialized on harvesting Twitter data of less frequent languages.
It searches continuously for new users that post tweets in the language of interest by querying the Twitter Search API 
for the most frequent and unique words in that language. Once a set of new potential users posting in the language 
of interest are identified, their full timeline is retrieved and the language identification is run over their timeline.
If it is evident that specific users post predominantly in the language of interest, they are added to the user 
list and their posts are being collected for the remainder of the collection period. 
In the case of Slovenian Twitter, the collection procedure started in August 2017 and is still running.
As a consequence, we are confident that the full Slovenian tweetosphere is well covered. 

\paragraph*{Data availability.}
All Twitter data were collected through the public Twitter API and
are subject to Twitter terms and conditions.
The Slovenian Twitter dataset 2018--2020 is available at a public language resource repository 
\textsc{clarin.si} at \url{https://hdl.handle.net/11356/1423}.

\paragraph*{Code availability.}
The code used to implement the Ensemble Louvain algorithm described in the paper is available 
at the Github repository \url{https://github.com/boevkoski/ensemble-louvain.git}.

\FloatBarrier
\subsection{Ensemble Louvain}
\label{sec:ensemble}

The Ensemble Louvain algorithm addresses the problem of instability of the
Louvain community detection algorithm. The instability is manifested by different results
of community detection in the same network, run with different initial seeds.
This is due to theoretical issues with modularity maximization, 
and to heuristic nature of an efficient implementation of the algorithm.

We address this instability problem with a new approach called Ensemble Louvain. 
The steps of the algorithm are as follows:
\begin{enumerate}
\item run several trials of Louvain on the same network,
\item built a new network where a pair of the original nodes is linked if their 
total co-membership across all the Louvain trials is above a given threshold (e.g., 90\%),
\item identify the disjoints sets which represent the resulting communities.
\end{enumerate}

More trials eventually lead to more stable partitioning (see Fig~\ref{fig:fig7}),
but increase the computation time. 
We found a reasonable trade-off between 50 and 500 trials, depending on the network size.

We are not the first to use ensembles for community detection.
A combination of several different algorithms to create a refined partitioning was proposed in~\cite{Chakraborty2020ensemble}.
Re-sampling methods with variations of the same network were used by~\cite{Dahlin2013ensemble}.
\cite{Lancichinetti2012consensus} create weighted consensus graphs and then detect communities in the consensus graph.

We measure the stability of Ensemble Louvain by Normalized Mutual Information (NMI) and Adjusted Rand Index (ARI). 
An initial comparison between the standard Louvain versus Ensemble Louvain is performed on three well-known datasets:
the Football network (115 nodes), the Email EU core (1005 nodes), and a Slovenian retweet network (3992 nodes).
100 separate experiment runs show that Ensemble Louvain yields significantly more stable results, 
especially on the larger networks, where the variation between possible solutions grows.

We measure the performance with respect to the ``ground truth'' for the Football and Email EU Core networks. 
The initial results (presented by the mean $\pm$ standard deviation of the scores) show a significant 
improvement of Ensemble Louvain over the standard Louvain:
\begin{itemize}
\item The Football network, \\
standard Louvain: NMI=$0.88 \pm 0.015$ and ARI=$0.78 \pm 0.041$, \\
Ensemble Louvain: NMI=$0.92 \pm 0.008$ and ARI=$0.89 \pm 0.019$.
\item The Email EU Core network, \\
standard Louvain: NMI=$0.58 \pm 0.016$ and ARI=$0.32 \pm 0.032$, \\
Ensemble Louvain: NMI=$0.72 \pm 0.005$ and ARI=$0.52 \pm 0.012$.
\end{itemize}

\FloatBarrier
\subsection{BCubed measure of community similarity}
\label{sec:bcubed}

The BCubed measure was originally proposed to evaluate effectiveness of document clustering~\cite{Bagga1998bcubed}.
Its properties were compared to a wide range of other extrinsic clustering evaluation metrics,
with the conclusion that BCubed satisfies all the required qualitative properties~\cite{Amigo2009bcubed}.
Since data clustering and community detection in networks produce analogous results,
one can also apply the BCubed measure to evaluate the detected communities.
Communities can be evaluated against the ``ground truth'' when available,
or compared to each other, as is the case with evolving communities.

The BCubed measure is applicable to individual nodes, communities, and network partitions in general.
It decomposes the evaluation into calculating the precision and recall associated with each node 
in the network. The precision ($\pre$) and recall ($\rec$) are then combined into the $\fone$ score:
$$
\fone = 2\,\frac{\pre \cdot \rec}{\pre + \rec}.
$$
The $\fone$ score is a special case of Van Rijsbergen's effectiveness measure~\cite{VanRijsbergen1979}, 
where precision and recall can be combined with different weights.
In the following we focus on definitions of precision and recall for different cases,
and assume a balanced definition of the $\fone$ score as the harmonic mean.
We first define the BCubed measure for a node, and then proceed with definitions of 
\textit{core-$\fone$}, \textit{standard $\fone$}, and theoretical \textit{max-$\fone$}.

Let $L(n)$ denote the ``ground truth'' community and $C(n)$ the detected community
of the node $n$, $n \in L(n), C(n)$.
$\pre$ and $\rec$ for a node are defined as follows:
$$
\pre(n) = \frac{\left|L(n)\cap C(n)\right|}{\left|C(n)\right|},
$$
$$
\rec(n) = \frac{\left|L(n)\cap C(n)\right|}{\left|L(n)\right|}.
$$

\paragraph*{\textit{Core-$\fone$}.}
Let first assume a special case when a pair of network partitions consist of the same
set of nodes. In this case, we name the BCubed measure \textit{core-$\fone$}.
Let $Ls = \{L_i\}$ denote a set of ``ground truth'' communities $L_i$, and
$Cs = \{C_i\}$ a set of detected communities $C_i$. Constituent $\pre$ and $\rec$
for the partition $Cs$ with respect to $Ls$ are defined as:
$$
\pre(Cs | Ls) = \frac{1}{|Cs|} \sum_{n \in C_i, C_i \in Cs} \pre(n),
$$
$$
\rec(Cs | Ls) = \frac{1}{|Ls|} \sum_{n \in L_i, L_i \in Ls} \rec(n).
$$
The $\fone$ measure proposed by Rossetti~\cite{Rossetti2016f1} is a special case
of the \textit{core-$\fone$}. In our case, the $\pre$ and $\rec$ are computed with respect to
all the communities $Ci$ and $Li$, while Rossetti computes the $\pre$ and $\rec$ just between
a pair of communities with the largest overlap.

\paragraph*{\textit{Standard $\fone$}.}
In general, a pair of partitions $P_0, P_1$ has some overlapping nodes, and some nodes
that are present in only one of the partitions. Let $Ls, Cs$ denote communities with overlapping nodes,
and $R_0, R_1$ the nodes specific to the respective partitions $P_0, P_1$. We have:
$$
P_0 = Ls \cup R_0 ,\;\;\; P_1 = Cs \cup R_1.
$$
$\pre$ and $\rec$ of partition $P_1$ with respect to the ``ground truth''
partition $P_0$ are then computed as follows:
$$
\pre(Cs | P_0) = \pre(Cs | Ls),
$$
$$
\pre(P_1 | P_0) = \frac{|Cs|}{|Cs|+|R_1|} \pre(Cs | Ls),
$$
$$
\rec(Cs | P_0) = \frac{|Ls|}{|Ls|+|R_0|} \rec(Cs | Ls),
$$
$$
\rec(P_1 | P_0) = \rec(Cs | P_0).
$$

\paragraph*{\textit{Max-$\fone$}.}
A theoretical maximum value of $\fone$ can be computed under the assumption that all the
overlapping nodes of the two partitions $P_0, P_1$ form one community. Let $C = L$ denote
the community with the intersecting nodes, $R_0$ extra nodes in $P_0$ (w.r.t. $P_1$), and
$R_1$ extra nodes in $P_1$ (w.r.t. $P_0$):
$$
C = L = P_1 \cap P_0, \;\;\; P_0 = L \cup R_0 ,\;\;\; P_1 = C \cup R_1.
$$
$\pre$ and $\rec$ of $P_1$ with respect to the ``ground truth'' $P_0$ are computed as:
$$
\pre(P_1 | P_0) = \frac{|C|}{|C| + |R_1|} = \frac{|C|}{|P_1|},
$$
$$
\rec(P_1 | P_0) = \frac{|C|}{|L| + |R_0|} = \frac{|C|}{|P_0|}.
$$
The \textit{max-$\fone$} score is then:
$$
\fone(P_1 | P_0) = 2\,\frac{\pre(P_1 | P_0) \cdot \rec(P_1 | P_0)}{\pre(P_1 | P_0) + \rec(P_1 | P_0)} 
= 2\,\frac{|C|}{|P_1| + |P_0|} 
= 2\,\frac{|P_1 \cap P_0|}{|P_1| + |P_0|}.
$$
This measure of similarity of two sets, $P_0$ and $P_1$, is also known as S\o{}rensen-Dice 
coefficient~\cite{Sorensen1948, Dice1945}. It is directly related to the Jaccard index:
$$
\jac(P_1 | P_0) = \frac{|P_1 \cap P_0|}{|P_1 \cup P_0|}.
$$
The transformation between the Jaccard index and $\fone$ is as follows:
$$
\jac = \frac{\fone}{2 - \fone}, \;\;\; \fone = \frac{2 \cdot \jac}{1 + \jac}.
$$

The BCubed-based $\fone$ measure therefore has two special cases, 
\textit{core-$\fone$} for comparing completely overlapping network partitions,
and \textit{max-$\fone$} for comparing two partitions with emerging (new) and
disappearing (lost) nodes. The later case is specially relevant in evolving retweet networks,
when new users appear and some users leave the network at different time windows.

\begin{figure*}[!ht]
\begin{center}
\includegraphics[width=\textwidth]{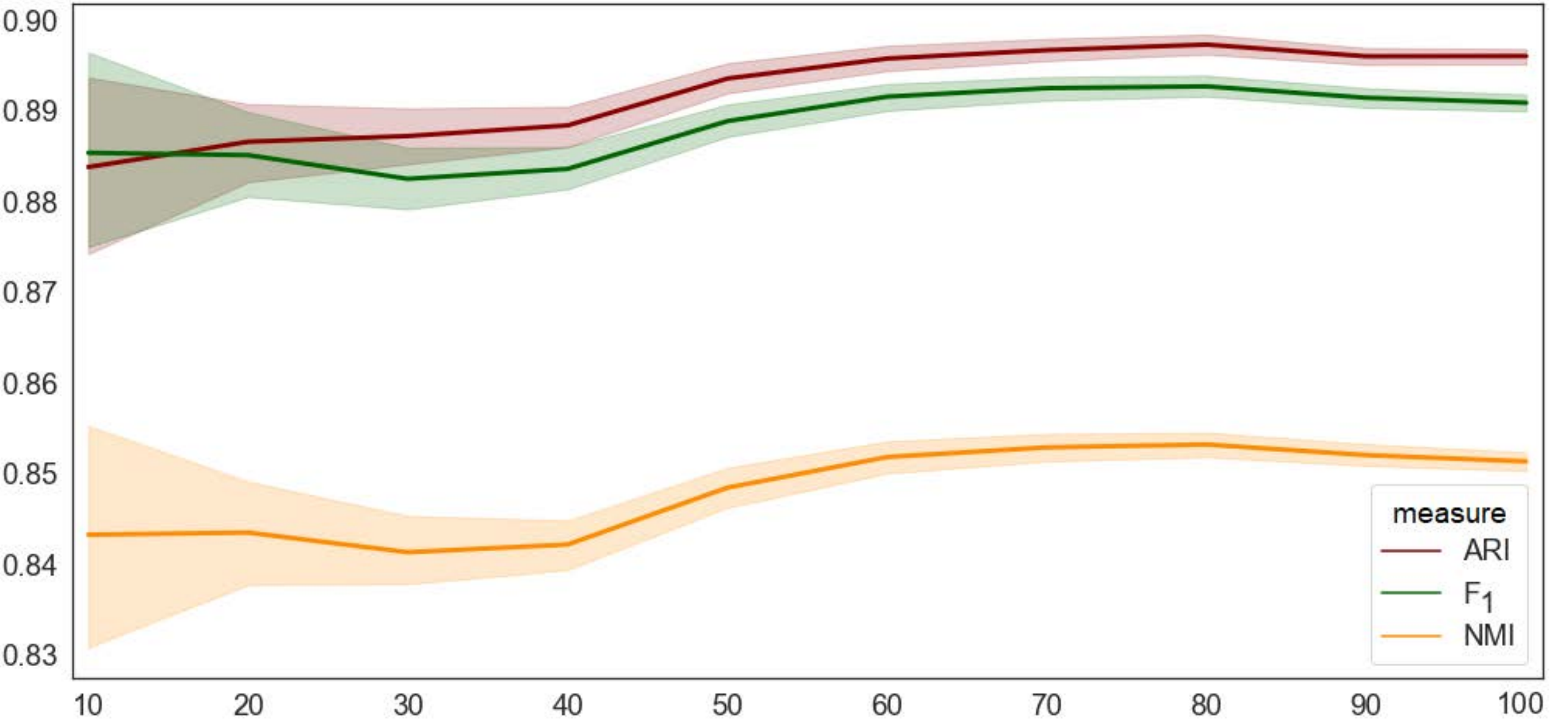}
\caption{
\textbf{A comparison of the BCubed $\fone$ measure with Adjusted Rand Index (ARI) and
Normalized Mutual Information (NMI).}
The comparison is run on the initial $G_0$ 
Slovenian retweet network. On the x-axis is the number of standard Louvain trials,
$N = 10, 20, \ldots, 100$. For each $N$, all the resulting partitions are pairwise compared by the 
three measures, ARI, $\fone$, and NMI (y-axis). Solid lines show the mean values 
and shaded areas the 95\% confidence intervals.
}
\label{fig:fig7}
\end{center}
\end{figure*}

The $\fone$ can be compared to standard community evaluation measures, such as
Adjusted Rand Index (ARI) and Normalized Mutual Information (NMI).
In Fig~\ref{fig:fig7} we compare the three measures on the same network,
running several trials of the standard Louvain with different initial seeds.
The $\fone$ in this case is actually the \textit{core-$\fone$}, compatible to ARI and NMI.
ARI and NMI cannot be applied in the case when network partitions differ in
the sets of respective nodes.

\FloatBarrier
\section*{Acknowledgments}

The authors acknowledge financial support from the Slovenian Research Agency (research core 
funding no. P2-103 and P6-0411), the Slovenian Research Agency and the 
Flemish Research Foundation bilateral research project LiLaH (grant no. ARRS-N6-0099 
and FWO-G070619N), and the European Union's Rights, Equality and Citizenship Programme 
(2014-2020) project IMSyPP (grant no. 875263). The European Commission's 
support for the production of this publication does not constitute an 
endorsement of the contents, which reflect the views only of the authors, 
and the Commission cannot be held responsible for any use which may be made 
of the information contained therein.

\FloatBarrier
\nolinenumbers

\begin{thebibliography}{10}

\bibitem{Dakiche2019tracking}
Dakiche N, Tayeb FBS, Slimani Y, Benatchba K.
\newblock Tracking community evolution in social networks: A survey.
\newblock Information Processing \& Management. 2019;56(3):1084--1102.

\bibitem{Evkoski2020polarization}
Evkoski B, Mozeti\v{c} I, Ljube\v{s}i\'{c} N, Novak PK.
\newblock Evolution of political polarization on {Slovenian Twitter}.
\newblock In: Complex Networks 2020, Book of abstracts; 2020. p. 325--327.

\bibitem{Rossetti2018}
Rossetti G, Cazabet R.
\newblock Community discovery in dynamic networks.
\newblock ACM Computing Surveys. 2018;51(2):1--37.
\newblock doi:{10.1145/3172867}.

\bibitem{coscia2011classification}
Coscia M, Giannotti F, Pedreschi D.
\newblock A classification for community discovery methods in complex networks.
\newblock Statistical Analysis and Data Mining: The ASA Data Science Journal.
  2011;4(5):512--546.

\bibitem{holme2012temporal}
Holme P, Saram{\"a}ki J.
\newblock Temporal networks.
\newblock Physics reports. 2012;519(3):97--125.

\bibitem{aynaud2013communities}
Aynaud T, Fleury E, Guillaume JL, Wang Q.
\newblock Communities in evolving networks: definitions, detection, and
  analysis techniques.
\newblock In: Dynamics On and Of Complex Networks, Volume 2. Springer; 2013. p.
  159--200.

\bibitem{hartmann2016clustering}
Hartmann T, Kappes A, Wagner D.
\newblock Clustering evolving networks.
\newblock In: Algorithm engineering. Springer; 2016. p. 280--329.

\bibitem{lambiotte2016guide}
Lambiotte R, Masuda N.
\newblock A guide to temporal networks. vol.~4.
\newblock World Scientific; 2016.

\bibitem{chen2010detecting}
Chen Z, Wilson KA, Jin Y, Hendrix W, Samatova NF.
\newblock Detecting and tracking community dynamics in evolutionary networks.
\newblock In: 2010 IEEE International Conference on Data Mining Workshops.
  IEEE; 2010. p. 318--327.

\bibitem{bota2010community}
B{\'o}ta A, Csizmadia L, Pluh{\'a}r A.
\newblock Community detection and its use in real graphs.
\newblock Matcos. 2010;.

\bibitem{alvari2014community}
Alvari H, Hajibagheri A, Sukthankar G.
\newblock Community detection in dynamic social networks: A game-theoretic
  approach.
\newblock In: 2014 IEEE/ACM International Conference on Advances in Social
  Networks Analysis and Mining (ASONAM). IEEE; 2014. p. 101--107.

\bibitem{agarwal2012real}
Agarwal MK, Ramamritham K, Bhide M.
\newblock Real time discovery of dense clusters in highly dynamic graphs:
  Identifying real world events in highly dynamic environments.
\newblock In: Proc. VLDB. vol.~5; 2012. p. 980--991.

\bibitem{crane2015community}
Crane H, Dempsey W. Community detection for interaction networks; 2015.
\newblock Available from: \url{https://arxiv.org/abs/1509.09254}.

\bibitem{aynaud2011multi}
Aynaud T, Guillaume JL.
\newblock Multi-step community detection and hierarchical time segmentation in
  evolving networks.
\newblock In: Proc. 5th SNA-KDD workshop; 2011. p. 69--103.

\bibitem{gauvin2014detecting}
Gauvin L, Panisson A, Cattuto C.
\newblock Detecting the community structure and activity patterns of temporal
  networks: a non-negative tensor factorization approach.
\newblock PLoS ONE. 2014;9(1):e86028.

\bibitem{aynaud2010static}
Aynaud T, Guillaume JL.
\newblock Static community detection algorithms for evolving networks.
\newblock In: 8th International symposium on modeling and optimization in
  mobile, ad hoc, and wireless networks. IEEE; 2010. p. 513--519.

\bibitem{Cherepnalkoski2016retweet}
Cherepnalkoski D, Mozeti\v{c} I.
\newblock Retweet networks of the {European Parliament}: Evaluation of the
  community structure.
\newblock Applied Network Science. 2016;1(1):2.
\newblock doi:{10.1007/s41109-016-0001-4}.

\bibitem{Cherepnalkoski2016cohesion}
Cherepnalkoski D, Karpf A, Mozeti\v{c} I, Gr\v{c}ar M.
\newblock Cohesion and coalition formation in the {European Parliament}:
  Roll-call votes and {Twitter} activities.
\newblock PLoS ONE. 2016;11(11):e0166586.
\newblock doi:{10.1371/journal.pone.0166586}.

\bibitem{Grcar2017stance}
Gr\v{c}ar M, Cherepnalkoski D, Mozeti\v{c} I, {Kralj Novak} P.
\newblock Stance and influence of {Twitter} users regarding the {Brexit}
  referendum.
\newblock Computational Social Networks. 2017;4(1):6.
\newblock doi:{10.1186/s40649-017-0042-6}.

\bibitem{Fortunato2016community}
Fortunato S, Hric D.
\newblock Community detection in networks: A user guide.
\newblock Physics Reports. 2016;659:1--44.
\newblock doi:{10.1016/j.physrep.2016.09.002}.

\bibitem{Blondel2008fast}
Blondel VD, Guillaume JL, Lambiotte R, Lefebvre E.
\newblock Fast unfolding of communities in large networks.
\newblock Journal of Statistical Mechanics: Theory and Experiment.
  2008;2008(10):P10008.

\bibitem{Newman2006modularity}
Newman MEJ.
\newblock Modularity and community structure in networks.
\newblock Proceedings of the National Academy of Sciences.
  2006;103(23):8577--8582.

\bibitem{Lancichinetti2009community}
Lancichinetti A, Fortunato S.
\newblock Community detection algorithms: a comparative analysis.
\newblock Physical review E. 2009;80(5):056117.

\bibitem{Good2010modularity}
Good BH, de~Montjoye YA, Clauset A.
\newblock Performance of modularity maximization in practical contexts.
\newblock Phys Rev E. 2010;81:046106.
\newblock doi:{10.1103/PhysRevE.81.046106}.

\bibitem{Hubert1985ARI}
Hubert L, Arabie P.
\newblock Comparing partitions.
\newblock Journal of Classification. 1985;2(1):193--218.
\newblock doi:{10.1007/BF01908075}.

\bibitem{Danon2005NMI}
Danon L, D{\'{\i}}az-Guilera A, Duch J, Arenas A.
\newblock Comparing community structure identification.
\newblock Journal of Statistical Mechanics: Theory and Experiment.
  2005;2005(09):P09008--P09008.
\newblock doi:{10.1088/1742-5468/2005/09/p09008}.

\bibitem{Amigo2009bcubed}
Amig\'{o} E, Gonzalo J, Artiles J, Verdejo F.
\newblock A comparison of extrinsic clustering evaluation metrics based on
  formal constraints.
\newblock Information Retrieval. 2009;12(4):461--486.

\bibitem{Rossetti2016f1}
Rossetti G, Pappalardo L, Rinzivillo S.
\newblock A novel approach to evaluate community detection algorithms on ground
  truth.
\newblock In: 7th Workshop on Complex Networks; 2016.

\bibitem{Evkoski2021commhate}
Evkoski B, Pelicon A, Mozeti\v{c} I, Ljube\v{s}i\'{c} N, Novak PK. Retweet
  communities reveal the main sources of hate speech; 2021.
\newblock Available from: \url{https://arxiv.org/abs/2105.14898}.

\bibitem{Kralj2021hate}
Novak PK, Ljube\v{s}i\'{c} N, Pelicon A, Mozeti\v{c} I. Hate speech detection
  as a knowledge discovery process; 2021.

\bibitem{Sluban2015sentiment}
Sluban B, Smailovi\'{c} J, Battiston S, Mozeti\v{c} I.
\newblock Sentiment leaning of influential communities in social networks.
\newblock Computational Social Networks. 2015;2(1):9.
\newblock doi:{10.1186/s40649-015-0016-5}.

\bibitem{DelVicario2016echo}
Del~Vicario M, Vivaldo G, Bessi A, Zollo F, Scala A, Caldarelli G, et~al.
\newblock Echo chambers: Emotional contagion and group polarization on
  Facebook.
\newblock Scientific Reports. 2016;6(1):37825.
\newblock doi:{10.1038/srep37825}.

\bibitem{Hirsch2005index}
Hirsch JE.
\newblock An index to quantify an individual's scientific research output.
\newblock Proceedings of the National Academy of Sciences.
  2005;102(46):16569--16572.

\bibitem{Gallagher2021SustainedOA}
Gallagher RJ, Doroshenko L, Shugars S, Lazer D, Welles BF.
\newblock Sustained online amplification of COVID-19 elites in the United
  States.
\newblock Social Media + Society. 2021;7(2):20563051211024957.

\bibitem{Ljubesic2014tweetcat}
Ljube{\v{s}}i{\'c} N, Fi{\v{s}}er D, Erjavec T.
\newblock {T}weet{C}a{T}: {A} tool for building {T}witter corpora of smaller
  languages.
\newblock In: Proc. 9th Intl. Conf. on Language Resources and Evaluation.
  European Language Resources Association (ELRA); 2014. p. 2279--2283.

\bibitem{Chakraborty2020ensemble}
Chakraborty T, Park N, Agarwal A, Subrahmanian VS.
\newblock Ensemble detection and analysis of communities in complex networks.
\newblock ACM/IMS Transactions on Data Science. 2020;1.
\newblock doi:{10.1145/3313374}.

\bibitem{Dahlin2013ensemble}
Dahlin J, Svenson P. Ensemble approaches for improving community detection
  methods; 2013.
\newblock Available from: \url{https://arxiv.org/abs/1309.0242}.

\bibitem{Lancichinetti2012consensus}
Lancichinetti A, Fortunato S.
\newblock Consensus clustering in complex networks.
\newblock Scientific Reports. 2012;2(1):336.
\newblock doi:{10.1038/srep00336}.

\bibitem{Bagga1998bcubed}
Bagga A, Baldwin B.
\newblock Entity-based cross-document coreferencing Using the Vector Space
  Model.
\newblock In: Proc. 17th Intl. Conf. on Computational Linguistics (COLING).
  Stroudsburg, PA, USA; 1998. p. 79--85.

\bibitem{VanRijsbergen1979}
{Van Rijsbergen} CJ.
\newblock Information Retrieval.
\newblock 2nd ed. Newton, MA, USA: Butterworth; 1979.

\bibitem{Sorensen1948}
S\o{}rensen T.
\newblock A method of establishing groups of equal amplitude in plant sociology
  based on similarity of species and its application to analyses of the
  vegetation on Danish commons.
\newblock Kongelige Danske Videnskabernes Selskab. 1948;5(4):1--34.

\bibitem{Dice1945}
Dice LR.
\newblock Measures of the amount of ecologic association between species.
\newblock Ecology. 1945;26(3):297--302.
\newblock doi:{10.2307/1932409}.

\end{thebibliography}

%
%
%

\end{document}